# LBT Italia: Current Achievements and Future Directions

Silvia Tosi [1,*], Ester Marini [1], Felice Cusano [2], Andrea Rossi [2], Roberto Speziali [1] and Roberta Carini [1]

1   INAF, Osservatorio Astronomico di Roma, Via Frascati 33, 00077 Monte Porzio Catone, Italy; ester.marini@inaf.it (E.M.); roberto.speziali@inaf.it (R.S.); roberta.carini@inaf.it (R.C.)
2   INAF, Osservatorio di Astrofisica e Scienza dello Spazio, Via Piero Gobetti 93/3, 40129 Bologna, Italy; felice.cusano@inaf.it (F.C.); andrea.rossi@inaf.it (A.R.)
*   Correspondence: silvia.tosi@inaf.it

**Abstract**

The Large Binocular Telescope (LBT) is a world-leading astronomical observatory, where the Italian partnership has played an important role in increasing the telescope's productivity, both through an optimized observing strategy and through peer-reviewed publications that are well recognized by the international astronomical community. This manuscript provides an updated overview of the active and past instruments at LBT, together with key usage statistics. In particular, we analyze the operational performance recorded in the LBT Italia night logs during INAF's observing time and assess the scientific impact of each instrument. Between 2014 and 2025, LBT Italia produced an average of 14 refereed publications per year, based on an annual average of 311 h of on-sky time. This corresponds to approximately 2.2 nights of telescope time per publication. The results of this analysis are placed in an international context to evaluate the competitiveness of LBT, and we outline future perspectives for scientific exploitation.

## 1. Introduction

The LBT is a binocular telescope located at an elevation of 3,221 m on Mount Graham in Arizona, USA, as part of the Mount Graham International Observatory (MGIO). It consists of two 8.4 m primary mirrors mounted side by side, separated by 14.4 m center to center, offering an interferometric baseline of 22.8 m. This design, combined with state-of-the-art adaptive optics using Gregorian adaptive secondary mirrors, ensures a high angular resolution, a low thermal background, and exceptional sensitivity, which is ideal for detecting faint astronomical sources [1–4].

Leveraging these capabilities, the Large Binocular Telescope Observatory (LBTO) ranks as one the most powerful 8 m class telescopes, enabling a broad array of scientific investigations, from the study of exoplanets to observations of the high-redshift universe (e.g., [5,6]). Its innovative binocular design offers unique advantages, such as the ability to observe with both mirrors simultaneously or independently, as well as interferometric resolution, which support competitive and diverse observing programs. Furthermore, LBTO plays a pivotal role in bridging current-generation facilities and the upcoming era of Extremely Large Telescopes (ELTs), including the Giant Magellan Telescope (GMT, [7]) the Thirty Meter Telescope (TMT, [8]), and the European Extremely Large Telescope (EELT, [9]). Notably, the LBT serves as the first operational ELT-class observatory, enabling science at the 23 m scale and acting as a testbed for technologies essential to the next generation of astronomical instrumentation [10,11].

Regular scientific operations at the LBT began in 2006, while the current international collaboration between institutions in the United States, Italy, and Germany was formally established in 2010. Specifically, the current LBT Corporation includes

- Arizona Board of Regents (representing the Arizona State University, the Northern Arizona University, and the University of Arizona);
- The Istituto Nazionale di Astrofisica (INAF), Italy;
- The LBT Beteiligungsgesellschaft, Germany (on behalf of the Landessternwarte Heidelberg, Leibniz-Institut für Astrophysik Potsdam, Max-Planck-Institut für Astronomie, Max-Planck-Institut für Extraterrestr. Physik, and Max-Planck-Institut für Radioastronomie);
- Ohio State University, the United States (representing Ohio State University, the University of Notre Dame, the University of Minnesota, and the University of Virginia).

Among the LBT members, INAF holds 25% of the total observing time, which is managed by researchers of the LBT Italia group. Its mission is to coordinate Italian access to the telescope and to maximize the scientific return for the national astronomical community. In fulfilling this role, LBT Italia plays a crucial part in supporting Italian astronomers by ensuring the delivery of high-quality data products and by contributing significantly to the observatory's overall scientific output. Support to Principal Investigators (PIs) is provided throughout the entire lifecycle of their observing programs, from the call for proposals to the delivery of science-ready data.

In addition, the core responsibilities of LBT Italia include (1) scheduling observing programs according to their ranking, Moon phase, and sky conditions; (2) executing astronomical observations; and (3) providing support and tools for spectroscopic data reduction. LBT Italia also ensures that all reduced data from LBT facility instruments are made publicly available by archiving them in the LBT Reduced Science Data Center (LSC).

Given the complexity of the LBT's organizational structure and the continuous upgrades the telescope has undergone, this work aims to present the current status of the observatory and the key scientific achievements of LBT Italia from 2007 to 2025. It highlights LBT Italia's contributions to the international astronomical community and its role in advancing ground-based observational capabilities.

In Section 2, we describe the instruments that have been installed on the LBT, highlighting those that are currently operational. Section 3 focuses on how the observing time allocated to LBT Italia is utilized, including an evaluation of time lost due to weather conditions or technical issues. Section 4 presents the scientific impact of LBT Italia's observations based on the published research, and Section 5 places them in the context of 8–10 m class observatories to assess their significance. We conclude with a discussion (Section 6) summarizing recent achievements and future prospects for LBT Italia.

## 2. Instrumentation

The unique configuration of the LBT allows several instruments to be installed in nearly identical pairs, with one mounted onto each side of the telescope. Thanks to its binocular design, the LBT can support instruments that are optimized for different wavelength ranges but share the same observing mode. An example are the two SHARK instruments with adaptive optics (see Section 2.6). In other configurations, both sides of the telescope can feed a single instrument, such as PEPSI (see Section 2.4). This flexibility is not available for all instruments, including those designed for interferometry, such as LBTI (see Section 2.5). However, for the instruments that support it, the LBT offers three distinct binocular observing modes:

(a) Twin Mode: Both instruments are configured identically, allowing the same observation to be performed simultaneously with both mirrors. Using both mirrors in twin mode doubles the collecting area, allowing the same depth to be reached in about half the exposure time under photon-limited conditions.

(b) Fraternal Mode: Instruments operate in parallel but with different configurations, such as different filters or masks, enabling two different observations to be conducted simultaneously.

(c) Hybrid Binocular Mode: Two different instruments are used on each side of the telescope, allowing for simultaneous collection of complementary data, such as photometric and spectroscopic observations.

In what follows, we refer to each pair as a single instrument for simplicity, while specific differences between the two channels (when relevant) are described in the dedicated sections. A schematic view showing the approximate locations of all currently active instruments at the LBT is presented in Figure 1.

- Facility Instruments: These include the Large Binocular Camera (LBC), the Multi-Object Double Spectrograph (MODS), the LBT Utility Camera in the Infrared (LUCI), and the Potsdam Echelle and Polarimetric Spectroscopic Instrument (PEPSI). These instruments were developed by the LBT members and subsequently handed over entirely to LBTO for operation. Observing time with these binocular instruments is managed by each partner through the classical mode, meaning that observations are conducted at the telescope during the time blocks allocated annually to each partner. In the case of the Italian partnership, observations with the facility instruments are performed either by a team of on-site observers at the LBT or remotely by members of the LBT Italia group. Specifically, when the LBT Italia team conducts observations on site, LBTO staff provide limited assistance and intervene only when explicitly requested. For remote observations, LBTO staff offer limited support by executing the observing plan according to instructions provided by LBT Italia members. It is important to note, however, that core telescope operations such as presetting, guiding, and dome setup remain the responsibility of the LBTO staff. This procedure ensures adherence to the scientific priorities established by the INAF Time Allocation Committee (TAC) and full compliance with the observational constraints specified by the users.

- PI Instruments: The current PI instrument at the LBT is the System for coronagraphy with High-order Adaptive optics from R to K bands (SHARK). This is divided into two channels, in the near-infrared (IR [12]) and visible (VIS [13]) regions, both providing imaging and coronagraphic modes. They are operated by the instrument team, and LBT members can gain access to observing time through direct agreements with the PI and by exchanging observing time.

- Unavailable instruments are instruments that are currently inactive at the LBT but have played, or are expected to play, a key role in supporting specific scientific goals (e.g., LINC-NIRVANA [14,15] and iLocater [16,17], respectively).

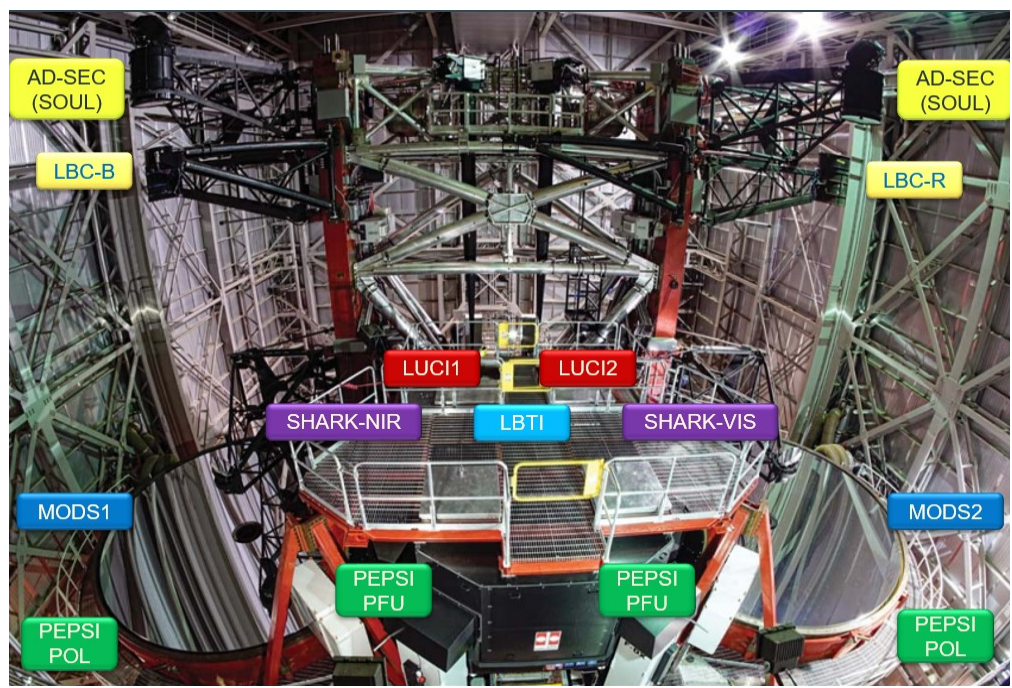

**Figure 1.** A photograph of the LBT with overlaid labels indicating the approximate locations of the instruments currently active at the telescope (credits: R. Cerisola).

*2.1. LBC*

The first instrument to be installed at the LBT was LBC [18,19], developed by INAF and mounted at the prime focus of the right-side telescope. First light with a single primary mirror and the LBC-Blue (LBC-B) prime-focus camera was achieved on 12 October 2005 (NGC 891). Binocular operation with the two Large Binocular Cameras (LBC-B and LBC-RED, hereafter LBC-R) has been in place since 2008. With the start of simultaneous observations on both channels in March 2008, the LBT officially entered its binocular era, making LBC the first binocular instrument installed at the observatory. Operating in this binocular mode, LBC supported the LBT's early scientific programs and remained the sole functioning instrument until the end of 2008, when the telescope's second instrument was installed.

The two LBC cameras are optimized for different spectral regions: LBC-B covers the ultraviolet and blue-optical bands (approximately 350–650 nm), while LBC-R is tailored to the red-optical regime (550–1000 nm). Their filter sets reflect these optimizations, with the r-Sloan filter being the only one shared between the two systems. Both channels offer a very wide field of view (FOV $\sim 23' \times 25'$), ideal for deep imaging and survey programs.

*2.2. LUCI*

The near-infrared spectrograph LUCI-1 was installed in September 2008 on one of the three Nasmyth rotators located on the left side of the bent Gregorian focal station. Its commissioning took place between September 2008 and November 2009 and was immediately followed by the start of science operations in December 2009. The second channel, LUCI-2, was installed in July 2013 and achieved first light on sky in January 2014. During 2011 and the early stages of LUCI-2 integration, LUCI-1 was temporarily dismounted from the telescope on two occasions to undergo upgrades. These interventions contributed to delaying the start of full binocular LUCI operations until semester 2017A.

The LUCI instruments [20] offer imaging and spectroscopy in the near-infrared, covering a wavelength range of approximately 0.9–2.5 μm over a $4' \times 4'$ field of view, and support multiple observing modes, including long-slit and multi-object spectroscopy (MOS, [21]). The spectroscopic resolving power reaches up to R $\sim$ 5800 using the 0.5″ slit, depending on the grating and wavelength setting, making it suitable for both survey work and detailed studies of specific targets.

The two LUCIs can be used either independently or simultaneously. When operated simultaneously, the pair supports both twin binocular mode (identical filters or gratings in each LUCI) and fraternal mode, in which the two arms observe the same field with different filters and/or gratings.

## 2.3. MODS

The first visible-light spectrograph, MODS-1, was installed at the direct Gregorian focus of the telescope in 2009 and began regular science operations in fall 2011. Its twin, MODS-2, achieved first light in February 2014 and joined MODS-1 for binocular science observations in spring 2015.

MODS [22] provides both imaging and spectroscopy in the optical range, covering approximately 320–1000 nm. It supports long-slit and multi-object spectroscopy. Light can either pass directly to the blue channel, be reflected toward the red channel, or be split by a dichroic at 565 nm, allowing for simultaneous observations in both channels. The dual-channel configuration enables simultaneous observations in the blue and red arms, enhancing spectral coverage and efficiency. When the two MODSs are run together, observers can operate MODS-1 and MODS-2 in an identical twin binocular setup or, in a fraternal configuration, assign different dispersers (e.g., the blue grating to MODS-1 and the red grating to MODS-2, or vice versa) to obtain simultaneous, complementary wavelength coverage. Each of the blue and red channels is equipped with medium-resolution gratings (R $\sim$ 2000) and low-resolution (R $\sim$ 150–500) double-pass prisms. Further specific details on MODS can be found in Pogge et al. [22,23].

Beginning with semester 2014A, the astronomical community gained access to the hybrid binocular mode configuration, which enabled simultaneous observations using LBC-R on the right telescope arm in combination with either MODS-1 or LUCI-1 on the left arm. This innovative setup allowed for concurrent photometric and spectroscopic data acquisition. In subsequent years, the configuration was expanded to include LBC-B paired with either MODS-2 or LUCI-2, further enhancing the observatory's multiplexing capabilities and scientific versatility.

## 2.4. PEPSI

Since semester 2015B (September 2015), the high-resolution spectrograph PEPSI [24] has been available at the LBT. The spectrograph itself is installed on a $2 \times 6$ m optical bench located within the telescope pier, inside a chamber with stabilized pressure and temperature. PEPSI is fed by two permanently mounted fiber units (PFUs), one at each Gregorian focus of the LBT. Each PFU transmits light through optical fibers approximately 40 m in length to the stabilized spectrograph located in the telescope pier [25].

PEPSI provides ultra-high-resolution spectroscopy across the 383–907 nm wavelength range (covered in three exposures), offering resolving powers from R $\sim$ 50,000 to 250,000, depending on the selected fiber mode and image slicer. Since September 2017, PEPSI has supported both spectroscopic and spectropolarimetric observations, with the latter enabled by dedicated polarimeters installed at each telescope's Gregorian focus [26].

*2.5. Adaptive Optics*

Since the installation of its adaptive secondary mirrors, the INAF community has had access to adaptive optics (AO) systems, which are technologies designed to correct for atmospheric distortions and significantly enhance image sharpness and spatial resolution. AO has become a central component of the LBT.

The first AO capability at the LBT was provided by the First Light Adaptive Optics (FLAO) system, a Single-Conjugate Adaptive Optics (SCAO) system, which relies on Natural Guide Stars (NGSs) to measure and correct atmospheric turbulence [27]. The FLAO system was installed onto the left Nasmyth platform of the telescope and was initially commissioned with the near-infrared camera PISCES (Prototype Imaging Spectrograph for Coronagraphic Exoplanet Studies; [28,29]). This configuration delivered some of the first high-resolution imaging results at the LBT and was later operated at multiple locations on the telescope. PISCES was equipped with a 1024 × 1024 pixel detector sensitive to the 1–2.5 μm wavelength range, and its configuration, in combination with the FLAO system, was used during two dedicated observing runs in 2012 and 2013, totaling 13 nights of observations. The PISCES camera was later decommissioned to make room for newer instruments.

Subsequently, AO has been used extensively with the Large Binocular Telescope Interferometer (LBTI; [30]), which enabled the first high-angular-resolution observations at the LBT using both imaging and nulling interferometry. Installed at the central bent Gregorian focus of the telescope, LBTI operates in the mid-infrared and combines light from the two 8.4 m primary mirrors through a cryogenic beam combiner. The system includes both slow alignment mechanisms and fast tip-tilt correction to compensate for atmospheric phase fluctuations [30,31]. The combined beam feeds the Nulling and Imaging Camera (NIC; [32]), which integrates three instruments: an L/M-band InfraRed Camera (LMIRcam; [5,33]), delivering diffraction-limited 1–5 μm imaging and spectroscopy over an 11″ field for exoplanet, disk, and zodiacal dust studies [34]; PhaseCam [35], a 2–2.4 μm sensor providing real-time tip-tilt and phase corrections, achieving the angular resolution of a 22.8 m telescope with the sensitivity of an 11.8 m mirror; and the Nulling Optimized Mid-Infrared Camera (NOMIC; [36,37]), an 8–13 μm camera optimized for exozodiacal light detection, supporting nulling interferometry, imaging, spectroscopy, and Fizeau mode with resolutions from 0.27″ (single aperture) to 0.10″ (dual aperture).

In 2018A, the Advanced Rayleigh Guided Ground Layer Adaptive Optics System (ARGOS; [38,39]) became available to LBT members. ARGOS employs four high-power lasers to correct turbulence in the lower layers of the atmosphere, a technique known as ground-layer correction. When paired with the LUCI N3.75 camera, ARGOS delivers a twofold improvement in seeing over a 4′ × 4′ field of view, achieving a full width at half maximum (FWHM) of just 0.2″ in the K band [40]. Although ARGOS underwent over 100 days of commissioning and was briefly offered in service blocks managed by the LBTO, it is currently not available for science operations.

Until the 2014 Call for Proposals for Instrument Upgrades and New Instruments, the FLAO systems operating at the LBT consisted of an adaptive secondary mirror (ASM) with 672 actuators and a Pyramid Wavefront Sensor (PWFS) with 30 × 30 sub-apertures. As a result, LUCI-2 achieved its first adaptive-optics-assisted observations in early 2015, followed by LUCI-1 in March 2016. This FLAO configuration was significantly enhanced by the SOUL (Single-Conjugated Adaptive Optics Upgrade for the LBT) project [41,42], with commissioning beginning on LUCI-1 in September 2018 [10]. SOUL introduced shorter readout times, reduced noise, and a larger detector format (240 × 240 pixels, 24 μm pixel size). These improvements enabled higher spatial sampling and allowed the system to approach the theoretical correction limit set by the photon flux of natural guide stars. As a result, the upgraded system achieved a diffraction-limited performance in the near-infrared using guide stars with R-band magnitudes as faint as 16.

In the following years, AO capability has been extended to the new SHARK-NIR and SHARK-VIS instruments (see Section 2.6 subsequently), both of which are now producing their first science results. In the near future, AO will also be fundamental for the high-resolution spectrograph iLocater and for the interferometric instrument LINC-NIRVANA, which exploits multi-conjugate AO to deliver wide-field correction (more details are reported in Section 2.7).

*2.6. SHARK-VIS and SHARK-NIR*

SHARK-VIS and SHARK-NIR are two instruments optimized for the direct imaging of exoplanets and circumstellar environments. They employ coronagraphic techniques and adaptive optics correction to achieve a diffraction-limited performance (see Section 2.5). Together, they provide two complementary channels for high-contrast, high-resolution imaging:

(a) SHARK-VIS began operations in October 2023, delivering data in the visible range ($\lambda \sim 0.4$–$0.9$ μm) on the right-side telescope. Its capabilities were previously tested through a forerunner experiment conducted in 2014 and 2015 [13].
(b) SHARK-NIR started scientific operations in 2024 and operates in the near-infrared range ($\lambda \sim 1$–$1.8$ μm) on the left-side telescope.

*2.7. Unavailable Instruments*

Throughout the LBT's operational years, various instruments have been developed and deployed both to enable pioneering scientific investigations and to serve as testbeds for advanced technologies. An example is LINC-NIRVANA, a near-infrared imager equipped with a layer-oriented multi-conjugate adaptive optics (MCAO) module and designed with a pathway to Fizeau interferometry. It achieved first light in April 2018 and subsequently underwent commissioning at the LBT. Designed to combine the light from the telescope's two 8.4 m mirrors, LINC-NIRVANA aimed to deliver an extremely high angular resolution (down to 10 mas) for detailed studies of compact astronomical sources. However, it is no longer in regular scientific use.

Looking to the future, iLocater is a diffraction-limited Doppler spectrometer currently in development for the LBT. It is characterized by a high spectral resolution (a median value of $R \sim 190{,}000$) and operates in the Y-band, used to characterize Earth-like exoplanets orbiting M-dwarf stars [43]. It employs the LBT's advanced adaptive optics system to correct for atmospheric turbulence, and it is designed to achieve radial velocity measurements with a sub-meter-per-second accuracy. The acquisition system for the left side of the telescope was installed in 2019 and successfully completed its commissioning phase. In May 2024, a prototype system known as "Little iLocater" [44] was temporarily deployed at the LBT to test the acquisition and fiber-injection performance, further validating the instrument's design. On-sky testing is currently expected to take place in semester 2025B.

## 3. Operational Performance of LBT Italia

To produce a detailed report on the main achievements of the Italian community at the LBT, we analyzed data collected from 2010 to 2024, including all available observations and log entries recorded during INAF's observing time. The statistics presented here include only programs executed through the INAF service or queue-scheduled time at the LBT and do not account for any collaborative programs involving pooled time from multiple partner institutions. Observatory-level initiatives (such as the PEPSI Exoplanet Transit Survey; see Keles et al. [45] and the references therein), as well as observations conducted outside the INAF queue (e.g., LBTI/LMIRcam campaigns), are not included in this analysis. Consequently, their associated observing time, completion rates, and scientific output are excluded from the efficiency and impact statistics presented in this work.

To properly interpret these statistics, it is important to consider the time allocation framework. Over the 15-year period examined, INAF was allocated an average of 547 h of observational time per year, equivalent to approximately 54 nights per year. Variations in the number of nights assigned to INAF and other partners are primarily due to the time allocated for instrument commissioning. For example, in 2010 and 2011, INAF received fewer nights (approximately 30–35 per year) due to the commissioning activities for LUCI-1 and MODS-1.

Throughout the analyzed period (2010–2024), LBT Italia ensured an optimized observing strategy by scheduling observations according to the program ranking, Moon phase, and current sky conditions. The schedule was regularly adjusted to optimize the observing efficiency, selecting programs most suitable for the prevailing conditions, which included seeing, cloud coverage, humidity, wind speed, and direction. Typical site conditions and forecast data for the period considered in this manuscript are available from the Advanced LBT Turbulence and Atmosphere (ALTA) Center, a project that provides automated nightly forecasts of optical turbulence and other astroclimatic and atmospheric parameters, especially those relevant to adaptive optics operations at the LBT (http://alta.arcetri.inaf.it/climatology/climatology.php, accessed on 31 July 2025). For example, the seeing is around $1''$ on more than 50% of the nights, with typical values ranging between $0.6''$ and $1.5''$. This result is based on analyses carried out by the ALTA Center, which report that the first seeing tertile is at $0.9''$, the second at $1.0''$, and the third at $1.4''$ (http://alta.arcetri.inaf.it/climatology/dimm.php, accessed on 31 July 2025).

In the upper panel of Figure 2, we present key statistics illustrating how the time available to LBT Italia has been used, excluding the commissioning time of the various instruments from this analysis. The figure shows, in purple, the total time available for LBT Italia observations; in red, the time lost due to technical issues; in green, the time lost to adverse weather conditions; in orange, the time spent on sky for conducting observations (including pointing, collimation, and calibration); and in blue, the science exposure time, which represents the effective time spent collecting scientific data on targets (the science exposure time is calculated separately for each telescope side rather than as a binocular pair, to account for periods when one channel was not operational). In this diagram, as well as in all subsequent figures, the gray shaded area highlights the 2019–2021 period, which corresponds to the COVID-19 pandemic. During this time, observations were fully suspended for approximately three months, and only one call for proposals was issued across the entire three-year span. For this reason, data from 2019 to 2021 are grouped together, unlike the rest of the analysis, which is based on two-year statistics.

*3.1. Time Lost*

Several factors can limit the operability of observations. For this reason, in this section, we illustrate the main situations reported through the LBT Italia night logs that led to interruptions during observing sessions. To preserve the structure and ensure high-quality data, various conditions require the suspension of data acquisition and the closure of the dome. For instance, observations are halted when the humidity exceeds 95%, rain showers or thunderstorms occur within 30 miles, or wind speeds surpass 20 m/s (45 mph) with gusts exceeding 22 m/s (50 mph).

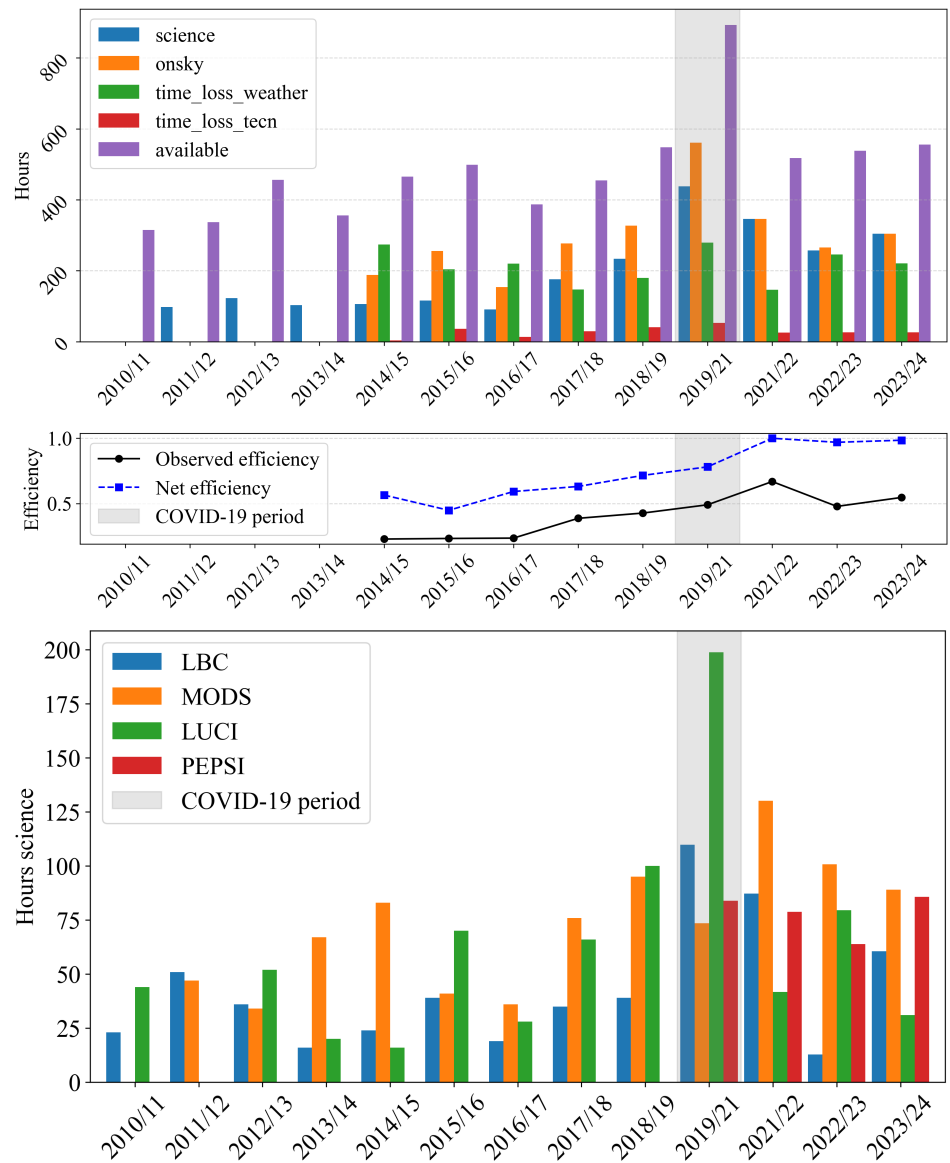

**Figure 2.** Upper panel: An annual overview of the operational time allocated to the Italian community at the LBT, showing the total available time (purple); losses due to technical issues (red) and adverse weather (green); on-sky time, including pointing, collimation, and calibration (orange); and science exposure time (blue), i.e., effective time spent collecting scientific data. Middle panel: Annual net and observed efficiency of LBT Italia observations, with observed efficiency defined as science exposure time over total available time and net efficiency as science exposure time over on-sky time (excluding technical and weather losses). Lower panel: The annual distribution of science exposure time among the current LBT Italia facility instruments. In all panels, the gray shaded area highlights the 2019–2021 period, corresponding to the COVID-19 pandemic.

These conditions, together with general safety protocols, result in an average loss of approximately 205 h per year, corresponding to 37% of the total available time, due to adverse weather conditions (green bars in the upper panel of Figure 2). In addition, about 31 h per year (6%) is lost due to technical issues affecting the telescope or instruments (red bars in the upper panel of Figure 2), which require prompt intervention and lead to the suspension of data acquisition. Losses due to human error are included within the category of technical time losses. Errors arising from incorrect setup or execution are grouped together with instrument and telescope malfunctions. This approach reflects the fact that within the LBT Italia time allocation, losses due to human error are significantly smaller than those caused by technical problems. Consequently, a separate record for human-related losses has not been established.

It is important to note that data for weather-related and technical losses are available only from the 2014/2015 to 2023/2024 observing seasons. Therefore, these values do not include time losses prior to 2014/2015. Additionally, the suspension of observations during the onset of the COVID-19 pandemic further contributes to the discrepancy between available time and science exposure time.

As a result, during the period between 2014 and 2024, the average effective observing time (i.e., on sky, orange bar in the upper panel of Figure 2) amounts to approximately 311 h per year, corresponding to about 31 nights of usable observations.

*3.2. Science Exposure Time*

For the portion of time available to LBT Italia that is not lost due to technical or weather-related issues, an average of 311 h per year is spent "on sky" (represented by the orange bars in the upper panel of Figure 2). Of this, approximately 248 h per year corresponds to science exposure on-target time. To evaluate operational performance, observed efficiency is defined as the ratio of science exposure time to the total available time, while net efficiency refers to the ratio of science exposure time to effective on-sky time (i.e., excluding weather and technical losses). Since science exposure is recorded separately for each side of the telescope, both efficiency metrics are calculated individually for each side. The total efficiency of the LBT can then be approximated as twice the single-side efficiency, with the latter shown in the middle panel of Figure 2.

Thanks to the dedicated efforts of on-site observers at the LBT, along with remote operators and members of the LBT Italia team, both the net and observed efficiency has improved over the years, highlighting optimization of the operational procedures and a reduction in technical downtime. Furthermore, as illustrated in the middle panel of Figure 2, both efficiency estimates focus on the period after 2014 to accurately account for time lost due to weather and technical interruptions.

To provide a more detailed breakdown, the lower panel of Figure 2 shows how the science exposure time has been distributed among the current facility instruments utilized by the Italian community (see Sections 2.1–2.4). To assess the distribution of science exposure time among the current facility instruments, we calculated the average number of hours per year for each instrument, considering only the years in which the instrument was actively used (i.e., excluding years with zero recorded exposure). This results in average yearly science exposure times of 42.5 h (17%) for LBC, 72.7 h (29%) for MODS, 62.3 h (24%) for LUCI, and 78.1 h (30%) for PEPSI.

*3.3. Time Requested Through LBT Italia Calls*

Every year, LBT Italia opens a call for proposals, receiving on average ∼44 submissions annually, which are then evaluated and ranked by the TAC. The scientific demand from the Italian community for the LBT's instruments exceeds the available observing time so that typically about 30% of the submitted proposals, usually those with the highest priority or with more flexible weather and seeing constraints, can be completed. This percentage does not account for the varying time requirements, which also influence the completion rate, as projects requesting more hours have a greater impact on the overall allocation. Nevertheless, it serves as a useful indicator of the high demand for LBT Italia observing time.

To maximize the number of concluded programs, LBT Italia implements optimized telescope operations, such as adjusting the night schedule based on weather conditions and instrument availability. Despite these efforts, the total observing time requested by the community generally exceeds the time allocated to LBT Italia programs (see Figure 3), and this discrepancy becomes even more critical when considering the actual science exposure time effectively available for observations (see Section 3.2). A notable exception occurred in 2020, when the COVID-19 pandemic stopped many scientific activities, leading to a reduction both in the number of observation hours requested and in those actually carried out by LBT Italia.

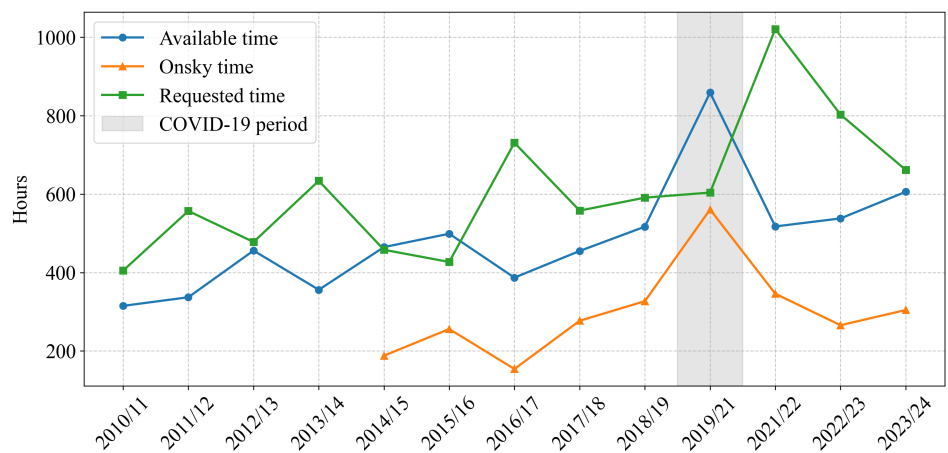

**Figure 3.** Total hours requested through LBT Italia proposals compared to the available observing time and actual on-sky hours each year.

## 4. Scientific Impact of LBT Italia Data

Most of the observations conducted by LBT Italia have resulted in high-quality data, leading to peer-reviewed publications in several prestigious scientific journals. A complete list of papers based on LBT data and used for the present analysis is available on the LBT Italia website (https://lbt.inaf.it/publications.php, accessed on 31 July 2025), which also includes papers produced during the commissioning of instruments conducted during LBT Italia's time. To ensure completeness and accuracy, this list was cross-checked with the information provided by the LBT Science Data Center (https://lsc.oa-roma.inaf.it/index.php, accessed on 31 July 2025) and the Astrophysics Data System (ADS) (https://ui.adsabs.harvard.edu/, accessed on 31 July 2025). According to this procedure, Figure 4 reports the number of publications per year for each of the instruments currently active at the LBT (blue: LBC; orange: MODS; green: LUCI; red: PEPSI; purple: LBTI; brown: SHARK-VIS; pink: SHARK-NIR) from 2007 to 2025. Since some papers are based on data from multiple instruments, the total number of publications per year is also reported (black line) to account for this overlap. Among the various instruments, LBC is the most frequently used in scientific publications, with an average of ∼5 papers per year, followed by LUCI and MODS, with ∼4 and ∼3 papers per year, respectively. PEPSI, SHARK-VIS, and SHARK-NIR contribute about one paper per year, while LBTI produces lower than one paper per year. Note that for the SHARK-VIS instrument, there are publications from 2017 related to a forerunner experiment, which was conducted to gather preliminary information (see Section 2.6).

We can conclude that the LBT's data leads to a substantial number of refereed publications, indicating a high level of scientific efficiency. In particular, considering the period between 2014 and 2025, LBT Italia produced an average of 14 refereed papers per year. With an annual average of 311 h of on-sky time, this corresponds to approximately 20.7 h of telescope time per publication.

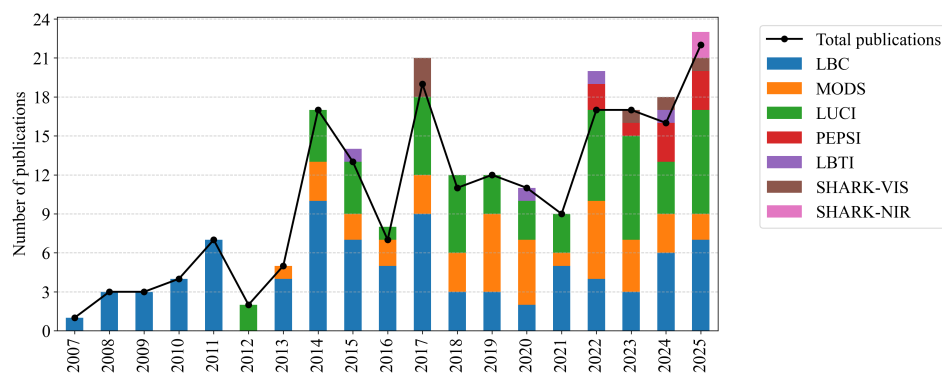

**Figure 4.** Number of publications per year for each currently active LBT instrument, shown as colored stacked bars. To account for papers referencing multiple instruments, the black line indicates the total annual number of publications based on data from these instruments.

## 5. LBT in the Context of 8–10 m Class Observatories

As anticipated in the previous section, the LBT's scientific contributions are well recognized by the international community, and its instrumentation is competitive with that of other leading facilities. For instance, the wide-field imaging capabilities of LBC are comparable to those of Subaru's HSC [46], while the LUCI and MODS spectrographs deliver a performance similar to that of VLT's X-shooter [47] and FORS2 [48], covering both the optical and near-infrared regimes. Notably, the LBT's high-resolution spectrograph PEPSI achieves resolving powers up to R∼250,000, exceeding what is typically available from instruments like Keck/HIRES [49] or GTC/MEGARA [50], especially in spectropolarimetric mode. In addition, like other major ground-based telescopes, the LBT makes use of adaptive optics; however, its system is integrated into the telescope architecture through adaptive secondary mirrors. This minimizes optical surfaces and optimizes the throughput, which is particularly valuable for high-contrast imaging applications such as those performed with SHARK.

Within the landscape of 8–10 m class telescopes, the LBT distinguishes itself through its binocular configuration of two co-mounted 8.4 m mirrors. This architecture differs from traditional monolithic designs and enables an interferometric baseline of 22.8 m, allowing the LBT to achieve an angular resolution comparable to that of the next generation of ELTs. In this way, the LBT already offers access to spatial resolutions that the astronomical community expects from much larger future facilities. Moreover, this unique configuration supports observing modes that are not available at other observatories. Its ability to operate in twin, fraternal, and hybrid binocular modes provides a real multiplexing advantage (see Section 2). For example, in hybrid mode, one can use instruments such as LBC and LUCI or MODS simultaneously for imaging and spectroscopy, increasing both flexibility and scientific return. Such synchronous, independently configurable apertures are not available at other telescopes in this class.

From the perspective of scientific efficiency, the LBT performs at a level comparable to that of other major facilities (see Section 4). For example, the VLT generated an average of approximately 587 refereed publications per year over the decade from 2014 to 2024, according to the ESO Telescope Bibliography (https://www.eso.org/sci/libraries/telbib_info.html, accessed on 31 July 2025). Assuming the output is approximately evenly distributed across the four Unit Telescopes and since around 90% of nights at Paranal are usable for science (https://www.eso.org/sci/facilities/paranal/astroclimate/Obsconditions.html, accessed on 31 July 2025), we estimate that about 21.8 h of observing time is required per publication. This value is very close to the 20.7 h per publication calculated for the LBT. Both observatories therefore require around 2.2 nights of telescope time per refereed paper. For these reasons, and despite the challenges associated with technical downtimes, the LBT remains competitive with the leading instruments at comparable observatories, achieving a scientific productivity fully in line with that of other major telescopes. The binocular configuration plays a key role in maintaining this level of efficiency, as it effectively increases the time spent on the target, even in cases where one side is temporarily unavailable due to maintenance or technical issues. Moreover, the publication and citation data discussed in Section 4 highlight a strong return on investment for the Italian community, confirming that the LBT serves not only as a complementary facility but also as a highly versatile and high-performing observatory.

## 6. Conclusions

Throughout its operational history, the LBT has collected a substantial volume of data that has been widely utilized by the scientific community to advance our understanding of the universe. Each year, LBT Italia receives approximately 44 proposals, with MODS being the most frequently employed instrument in Italian-led observations. The demand for observing time from the Italian community remains consistently high, resulting in a highly competitive environment where only the highest-ranked proposals, or those with more flexible observational constraints, are typically scheduled.

The data acquired through LBT Italia play a pivotal role in astronomical research, having so far contributed to 208 publications that have collectively received over 5400 citations. Specifically, between 2014 and 2025, LBT Italia produced an average of 14 refereed publications per year, based on an annual average of 311 h of on-sky time. This corresponds to approximately 2.2 nights of telescope time per publication, a level of productivity consistent with that of other major observatories, such as the VLT. Among these publications, the facility-class instruments represent the core of the Italian scientific output, while the newly operational PI instruments have already demonstrated strong promise and hold significant potential for future discoveries.

Over its years of operation, the LBT has already established itself as both a flagship 8 m class observatory and a technological bridge toward the era of ELTs. On a wider horizon, the LBT's binocular interferometric mode serves as a unique testbed for beam-combination techniques and phase-alignment strategies that are directly applicable to upcoming ELT-class facilities such as the CMT, TMT, and E-ELT, while its adaptive secondary mirrors continue to inform the next generation of adaptive optics modules for ELT instrumentation. Synergy with space missions will enhance the LBT's scientific impact further. For example, the LBT provides complementary ground-based imaging and spectroscopy for follow-up observations of transient events detected by missions like EUCLID [51,52], JWST [53], and the Roman Space Telescope [54], enabling more complete characterization of these phenomena. In addition, the LBT's role is strengthened further through continuous instrumental upgrades. For instance, during summer 2025, MODS-1 and MODS-2 underwent a controller upgrade (https://scienceops.lbto.org/, accessed on 31 July 2025), and new proposals are currently under evaluation following the recent call for new instrument concepts, which closed on 30 May 2025 (https://www.lbto.org/new-instrument-concepts/, accessed on 31 July 2025). In conclusion, the planned instrumental upgrades, operational refinements, and strategic positioning of the LBT guarantee that the Italian community will continue to deliver cutting-edge science and to pioneer techniques that will be central to the success of the upcoming ELT era.

**Funding:** This research received no external funding.

**Data Availability Statement:** The original contributions presented in this study are included in the article. Further inquiries can be directed to the corresponding author.

**Acknowledgments:** The authors acknowledge support from the INAF research project LBT—Supporto Arizona Italia.

**Conflicts of Interest:** The authors declare no conflicts of interest.